\documentclass{raa}
\usepackage{graphicx,times}
\usepackage{amssymb,amsmath}
\usepackage{rotating}

\usepackage{subfigure}
\usepackage{graphicx}
\usepackage{txfonts}
\usepackage{multirow}
\usepackage{subfigure}
\usepackage{supertabular}
\usepackage{lscape}

\usepackage[a4paper=true,dvipdfm=true,pagebackref=true]{hyperref}
\hypersetup{pdftitle = The title of my PDF, pdfauthor = My name, pdfsubject= The subject, pdfkeywords = keyword1 keyword2 keyword3}
\hypersetup{colorlinks = true, linkcolor = green, anchorcolor = red, citecolor = blue, filecolor = red, pagecolor = red, urlcolor = red}

\newcommand{\ha}{${\rm H}\alpha$}
\newcommand{\hb}{${\rm H}\beta$}

\newcommand{\nii}{{\sc [N ii]}}
\newcommand{\sii}{{\sc [S ii]}}

\newcommand{\oiii}{{\sc [O iii]}}
\newcommand{\kms}{$km\ s^{-1}$}

\begin{document}

   \title{A Search for Double-peaked narrow emission line Galaxies and AGNs in the LAMOST DR1
$^*$
\footnotetext{\small $*$ Supported by the National Natural Science Foundation of China and 973 Program.}
}

 \volnopage{ {\bf 2014} Vol.\ {\bf X} No. {\bf XX}, 000--000}
   \setcounter{page}{1}

   \author{
Zhi-Xin Shi\inst{1,2,3}
\and
A-Li Luo\inst{1}
\and
Georges Comte\inst{1,4}
\and
Xiao-Yan Chen\inst{1}
\and
Peng-Wei\inst{1,3}
\and
Yong-Heng Zhao\inst{1}
\and
Fu-Chao Wu\inst{2}
\and
Yan-Xia Zhang\inst{1}
\and
Shi-Yin Shen\inst{5}
\and
Ming Yang\inst{1}
\and
Hong Wu\inst{1}
\and
Xue-Bing Wu\inst{6}
\and
Hao-Tong Zhang\inst{1}
\and
Ya-Juan Lei\inst{1}
\and
Jian-Nan Zhang\inst{1}
\and
Ting-Gui Wang\inst{7}
\and
Ge Jin\inst{7}
\and
Yong Zhang\inst{8}
   }
%
   \institute{
Key Laboratory of Optical Astronomy, National Astronomical Observatories, Chinese Academy of Sciences, Beijing 100012, China; {\it lal@lamost.org}\\
\and%
National Laboratory of Pattern Recognition, Institute of Automation, Chinese Academy of Sciences, Beijing 100190, China\\
\and
University of Chinese Academy of Sciences, Beijing 100049, China\\
\and
Aix-Marseille Universit\'e \& CNRS, Institut Pyth\'eas, LAM (Laboratoire d'Astrophysique de Marseille), UMR7326, F-13388 Marseille, France\\
\and
Shanghai Astronomical Observatory, Chinese Academy of Sciences, Shanghai
200030, China;\\
\and
Department of Astronomy, Peking University, Beijing 100871, China;\\
\and
University of Science and Technology of China, Hefei 230026, China\\
\and
Nanjing Institute of Astronomical Optics \& Technology, National Astronomical Observatories, Chinese Academy of Sciences, Nanjing 210042, China\\
\vs \no
   {\small Received 2013 XX; accepted 2014 XX}
}

\abstract{
LAMOST has released more than two million spectra, which provide the opportunity to search for double-peaked narrow emission line (NEL) galaxies and AGNs.
The double-peaked narrow-line profiles can be well modeled by two velocity components, respectively  blueshifted and redshifted with respect to the systemic recession velocity.
This paper presents 20 double-peaked NEL galaxies and AGNs found from LAMOST DR1 using a search method based on multi-gaussian fit of the narrow emission lines.
Among them, 10 have already been published by other authors, either listed as genuine double-peaked NEL objects or as asymmetric NEL objects, the remaining 10 being first discoveries.
We discuss some possible origins for double-peaked narrow-line features, as interaction between jet and narrow line regions,
interaction with companion galaxies and black hole binaries.
Spatially resolved optical imaging and/or follow-up observations in other spectral bands are needed to further discuss the physical mechanisms at work.
}
   \authorrunning{Zhi-Xin Shi et al. }            
   \titlerunning{A Search for Double-peaked emission line Galaxies and AGNs in the LAMOST DR1}  
   \maketitle

%
\section{Introduction}           
\label{sect:intro}
The search for double-peaked narrow-line structure in galaxy spectra is an effective way of finding binary AGN candidates (Zhou et al. \cite{ref57}; Blecha et al. \cite{ref1}),
which are expected to take place in the final phases of the merging of two interacting active galaxies.
Since the suggestion that double-peaked profiles could be produced by binary AGN (Zhou et al. \cite{ref57}),
much attention has been paid to that hypothesis, yet few definite cases are known.
The most convincing examples can be found in  CXO J1426+35 (Barrows et al. \cite{ref25}), EGSD2 J1420+4259 (Gerke et al. \cite{ref2}),
NGC6240 (Komossa et al. \cite{ref59}), COSMOS J100043+020637 (Comerford et al. \cite{ref21}), SDSS J0952+2552 (McGurk et al. \cite{ref60}; Fu et al. \cite{ref26}) and so on
(see the review in Wang et al. \cite{ref58}, and Table 1 in Ge et al. \cite{ref9} ).
Several systematic searches for AGNs with double-peaked \oiii\ emission lines have been performed in the DEEP2 survey sample
(Gerke et al. \cite{ref2}; Comerford et al. \cite{ref3}) and in the Sloan Digital Sky Survey (hereafter SDSS) data releases
(Xu \& Komossa \cite{ref4}; Wang et al. \cite{ref5}; Liu et al. \cite{ref6, ref7}; Smith et al. \cite{ref8}; Ge et al. \cite{ref9}).
The last authors have retrieved a large sample of 3030 dual-peak NEL objects from the SDSS DR7, and a much larger sample of asymmetric profiles NEL objects, with in total 54 dual-cores candidates.
Among the candidates found at low or moderate redshift, when double-peaked \oiii\ emission line has been suspected or confirmed as indicator of a binary AGN,
the components most likely trace objects with spatial separations in the range from 100 pc to 10 kpc (Wang et al. \cite{ref5}).

However, the double-peaked NEL may also be produced by other mechanisms, such as chance superposition,
peculiar gas kinematics in the narrow-line regions (NLRs; e.g., Gelderman \& Whittle \cite{ref32}; Fu \& Stockton \cite{ref33}; Fischer
et al. \cite{ref34}; Shen et al. \cite{ref24}; Fu et al. \cite{ref26}) and
jet-cloud interactions (e.g., Stockton et al. \cite{ref35}; Rosario et al. \cite{ref36}). Because of the growing interest in double-peaked NEL objects, it is also important to check that the candidates are fully trustworthy, thus independant surveys should confirm this character.
In order to further analyse these double-peaked samples, follow-up observations are needed. Previous works have used various methods:
high-resolution optical imaging (Comerford et al. \cite{ref21}),
near-infrared imaging (Liu et al. \cite{ref6}; Fu et al.  \cite{ref22}; Rosario et al. \cite{ref23}; Shen et al. \cite{ref24}; Barrows et al. \cite{ref25}),
integral-field spectroscopy (Fu et al. \cite{ref26}),
hard X-ray observations (Comerford et al. \cite{ref27}; Civano et al. \cite{ref28}; Liu et al. \cite{ref29}),
radio observations (Zhou et al. \cite{ref57}; Fu et al. \cite{ref30}),
and long slit spectroscopy (Shen et al. \cite{ref24}; Comerford et al. \cite{ref31}).

In this paper, we focus on a systematic search for galaxies and AGNs with double-peaked NEL
in the First Data Release of the Sky Survey conducted with the Large Sky Area Multi-Object Fiber Spectroscopic Telescope (LAMOST, also called Guo Shou Jing Telescope, GSJT), hereafter LAMOST DR1.
Like the SDSS, LAMOST has produced a large number of quasar and galaxy spectra from October 2011 till June 2013 (its first two observing seasons). We have developed a method for searching double-peaked NEL galaxies and AGNs in LAMOST DR1,
and visually inspected the candidates spectra for confirmation. Our sample selection and method are described in Section 2.
We present and discuss the results in Section 3, and compare them to SDSS spectra derived results when available. We summarize this work  in Section 4.


\section{Sample selection}
\label{sect:prework}
\subsection{LAMOST}

LAMOST  has the capability of taking 4000 spectra of objects distributed
across a 25 square degrees field simultaneously in a single exposure.
The spectral range extends from 3700 to 9000 $\AA$ with a resolution of $R = 1800$. The telescope, fiber positioning system and its spectrographs are described
in detail in Cui et al. (\cite{ref10}).

A spectroscopic survey of over
10 million objects has started in autumn 2012, focusing on stellar astrophysics,
structure of the Milky Way, and extragalactic astrophysics and cosmology (Zhao et al. \cite{ref11}).
A limited Pilot Survey  dedicated to the test of data processing pipelines and exhaustive performance evaluation
was first launched on 2011 Oct 24, and ended in June 2012, including
nine lunar cycles and basically covering the best observing period available along a year at the Xinlong
Observing Station (Yao et al. \cite{ref12}; Luo et al. \cite{ref13}). In autumn 2012, LAMOST began a general survey.
The first public release, DR1 (July 2013), contains more than two million spectra.

The raw data are reduced with LAMOST 2D and 1D pipelines (Luo et al. \cite{ref14, ref13}), including bias subtraction,
cosmic-ray removal, spectrum extraction, flat-fielding through twilight exposures, wavelength calibration,
sky subtraction and multiple exposure coaddition. The LAMOST pipeline  only provides a relative flux calibration, hence, for the 17 objects for which SDSS spectra are available, we have used the SDSS fluxes  to rescale the relative LAMOST fluxes to the SDSS absolute flux scale.

\subsection{Method}
We start from the spectral sample of LAMOST DR1, including 12082
galaxies and 5017 quasars. In order to select double-peaked NEL
candidates, we have developed a reduction
method closely inspired from the work of Ge et al. \cite{ref9}.

\subsubsection{Sample trimming}
We first trim the DR1 extragalactic sample with the following criteria, respected by all preselected spectra:

(1) across the rest-frame wavelength ranges [4700, 5100] $\AA$  and [6500, 6800] $\AA$ (i.e. the regions containing respectively \hb - \oiii and \ha - \sii\ emission lines) the average  signal-to-noise ratio S/N per pixel of continuum must be $>$ 2;

(2) rest-frame equivalent widths of the emission lines must obey:
EW (\ha, \hb, \oiii\ $\lambda 5007$) $>$ 3 $\AA$;

(3) to ensure that \oiii\ emission lines are included in the observed
wavelength range, we initially cut the redshift at z $<$ 0.9.

This preselection provided a reduced sample of around 6000 galaxies and 3000 QSOs.
After visually inspecting each spectrum of LAMOST and selecting those susceptible to host double NEL,
we were left with about 150 galaxies and 50 QSOs,
which were then submitted to the following step.

\subsubsection{Starlight subtraction}
The spectra from LAMOST are taken through a 3.3 $\arcsec$ diameter
fiber, large enough to include the light from the nucleus
and  stellar light from the host galaxy. Absorptions can mask or
weaken the emissions, and the stellar continuum may also affect the
measurement of the emission-line intensities. To obtain reliable
values of emission-lines fluxes, the underlying stellar continuum is first removed using a spectral synthesis model.

\indent a) We first correct the foreground Galactic extinction using the reddening maps of Schlegel et al. (\cite{ref54})

\indent b) We shift back the spectra to the rest frame defined by a systemic redshift value automatically produced by
the LAMOST data processing pipeline. This redshift is fully consistent with the value attached to the objects in the SDSS, for 17 objects in common.

\indent c)The STARLIGHT software (Cid Fernandes et al. \cite{ref16})is then used to fit the continuum and stellar
absorption lines and continua from the underlying stellar population.
STARLIGHT generates a model spectrum $M_{\lambda}$ by co-adding
up to $N_{\ast}$ instantaneous starburst models (Simple Stellar Populations, SSPs) with different ages and
metallicities from the evolutionary synthesis models of Bruzual \& Charlot (\cite{ref15}).
This model is fitted to the observed spectrum $O_{\lambda}$ in which emission lines, night sky line residuals
and possible bad pixels have been masked out (see details in Chen et al. \cite{ref20}), using
the Metropolis scheme which searches for the minimum $\chi^{2}=\Sigma_{\lambda}[(O_{\lambda}-M_{\lambda})\omega_{\lambda}]^2$,
where $\omega_{\lambda}^{-1}$ is the error in $O_{\lambda}$ except for masked regions. Among the various options that can be selected,
 we use the reddening law of CCM (Cardelli et al. \cite{ref17}), the Padova 1994 tracks (Alongi et al. \cite{ref18}),
and the Chabrier (Chabrier \cite{ref19}) initial mass function.
Moreover, we use 45 SSPs, which included 15 different ages from 1 Myr to 13 Gyr and 3 different metallicities (0.2, 1, and 2.5 Z$_{\odot}$).

\begin{figure}
   \centering
   \includegraphics[width=0.8\textwidth, angle=0]{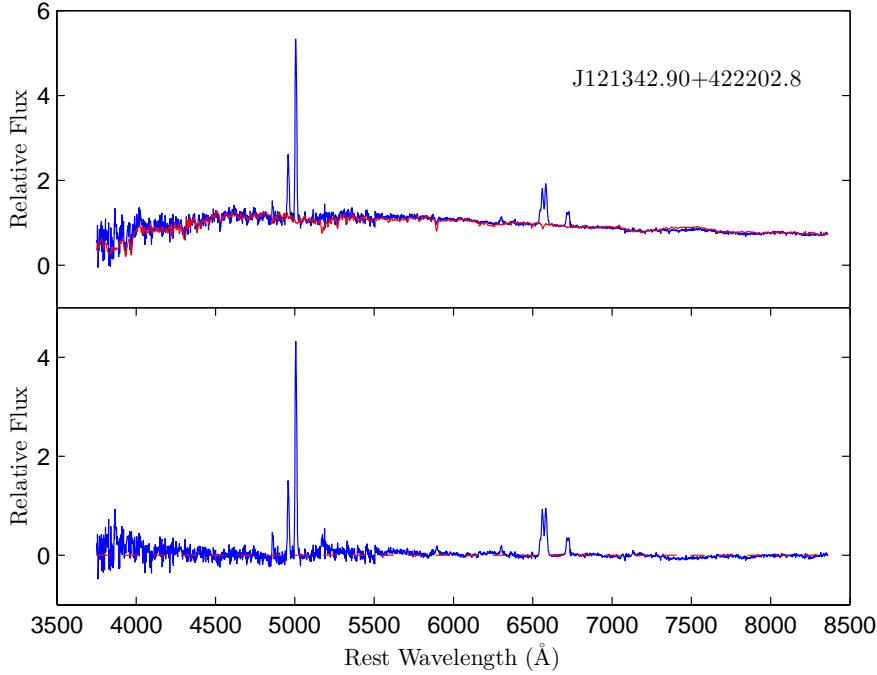}
   \caption{Example of host galaxy spectral fitting with stellar population templates (BC03, Bruzual \& Charlot (\cite{ref15}))  using the STARLIGHT code.
   In the top panel, the black line is the observed spectrum, the red line is the galaxy model spectrum. The bottom panel shows the residual, i. e., the star light subtracted spectrum.}
   \label{fig1-stel}
   \end{figure}

An example of starlight subtraction is shown in Fig. \ref{fig1-stel}, which exhibits both
the result of continuum and absorption lines fit (top panel) and the starlight subtracted spectrum (bottom panel). This latter is then submitted to the following procedure.

\subsubsection{Multi-gaussian fitting}
 We fit each individual detected emission line with three models: a single gaussian, a double gaussian and a double gaussian plus an additional gaussian broader wing .
The F-test is employed to decide which model has statistically the highest probability of being the best representation of the data. One calculates the F-value with the following equation:

 \begin{equation}
 F = \frac{(SSE_{m1}-SSE_{m2}) / (DoF_{m1}-DoF_{model2})}{SSE_{m2}/Dof_{m2}}
 \end{equation}

where $SSE_{m1}$ is the residual sum of squares for the simple gaussian model and $SSE_{m2}$ is for the complex gaussian model.
 $DoF$ is the number of degrees of freedom of each model (i.e. the number of data points used in the fit minus the number of function parameters used). A larger F-value indicates that the complex gaussian model has a highest probability to be correct than the single gaussian one.

 After we excluded the possibility of a single gaussian model, we use a free parameter set to fit each emission line,
and adopted them as the initial values for our fixed fitting technique. The ''blue'' (i.e. blueshifted from the adopted reference systemic redshift) and ''red'' (resp. redshifted) Gaussian systems show their respective consistency in having both  same center wavelength shift and same FWHM.

\begin{figure}
   \centering
   \includegraphics[width=\textwidth, angle=0]{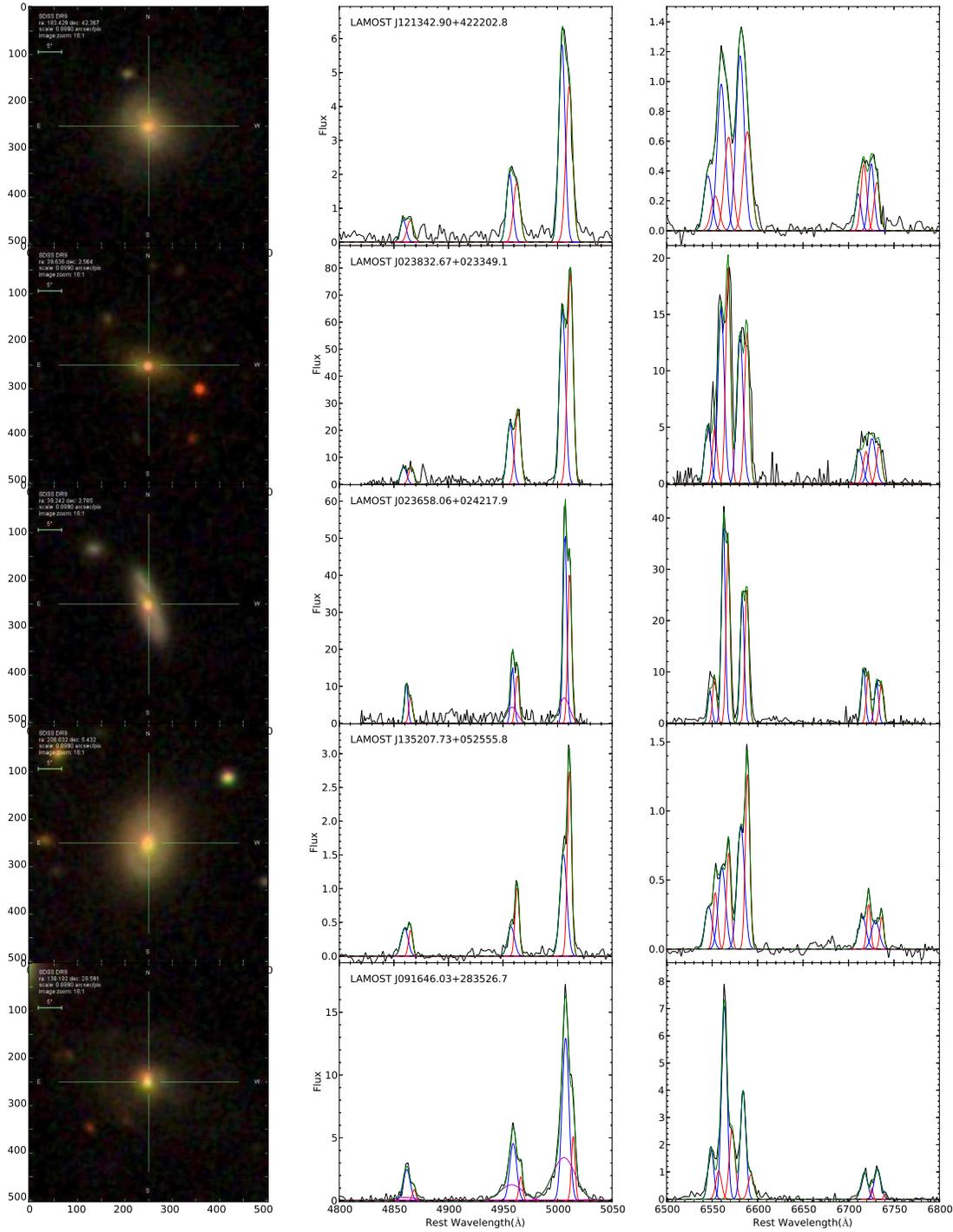}
   \caption{SDSS images (left) and spectra (right) of five examples with double peaked emission lines.
    Black: original spectra, blue: the blue component,
    red: the red component, magenta: the wing of emission lines,
    green: sum of all components. }
   \label{fig2-five}
   \end{figure}

 Finally, we visually inspect all the fits and verify that the line profiles are correctly fitted by the model.
 Then, we cross with the spectrum of SDSS to confirm the double NEL. There are 20 double-peaked NEL objects,
 and Fig. \ref{fig2-five} shows 5 examples from the galaxy subsample.

\begin{figure}
   \centering
   \includegraphics[width=0.8\textwidth, angle=0]{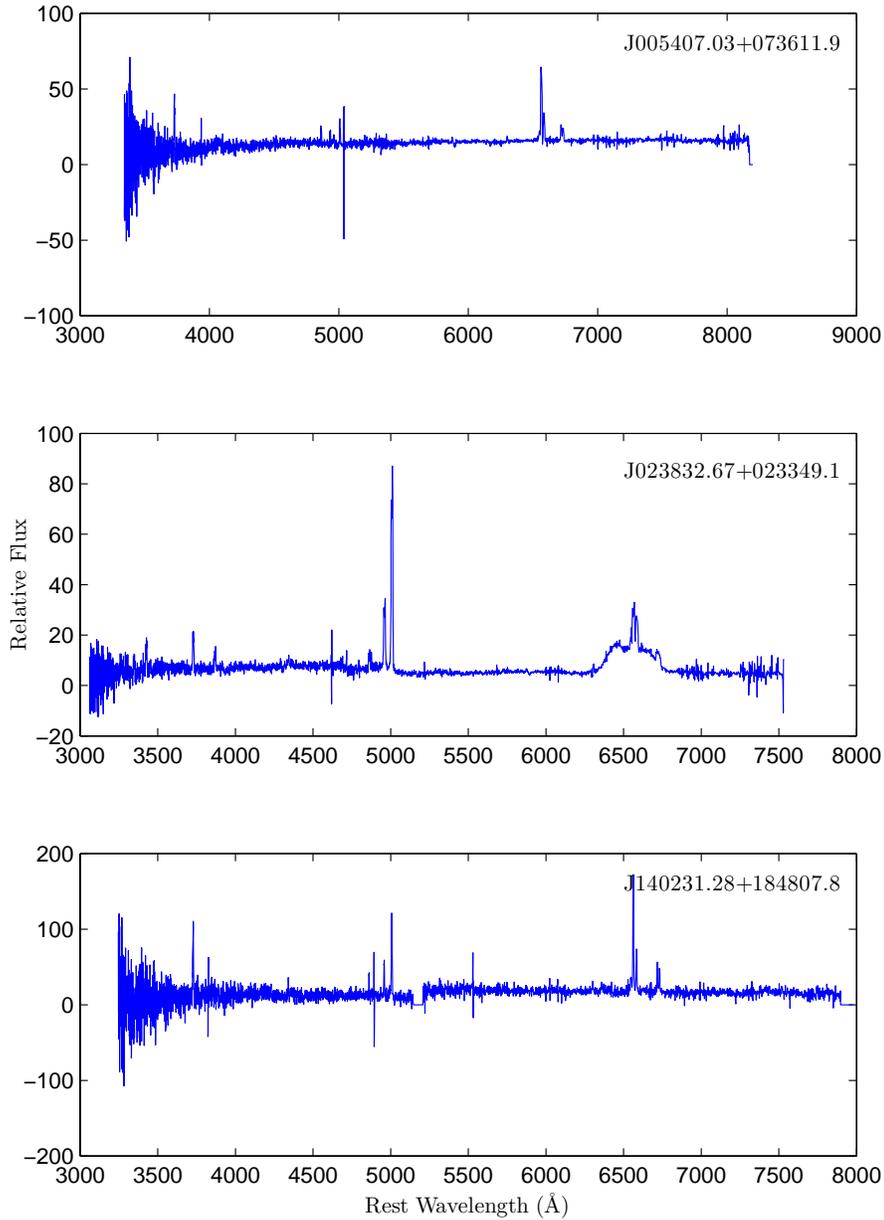}
   \caption{Totally, 20 double-peaked NEL objects were found from LAMOST DR1. Among 10 first discoveries, 3 new spectra were first observed by LAMOST and presented here.}
   \label{fig3-spectra}
   \end{figure}

\subsubsection{Comparison with SDSS spectra}

 In the present sample, objects detected as double-peaked NEL are located in the footprint of the SDSS. 17 have a spectrum in SDSS DR9 and 10 among these were already identified as double-peaked NEL or strongly asymmetric NEL (see Table \ref{tab2-emission} for references). We have collected the 17 SDSS spectra and have independently submitted them to the same procedure, which fully confirm the double-peaked nature of the narrow emission lines, ruling out possible noise or reduction artefact. We have also checked the consistency of the results derived from LAMOST and SDSS data on the velocity differences between the NEL components and on the emission line ratios of each NEL  component. We get:

\begin{equation}
\Delta V_{\rm LAMOST} = (0.94 \pm 0.10) \cdot \Delta V_{\rm SDSS} + (125 \pm 30)
\end{equation}
with $\Delta V$ in \kms, and
\begin{equation}
\log ([OIII]5007/H_{\beta})_{\rm LAMOST} = (0.91 \pm 0.05) \cdot \log  ([OIII]5007/H_{\beta})_{\rm SDSS} + (0.05 \pm 0.03)
\end{equation}
\begin{equation}
\log ([NII]6584/H_{\alpha})_{\rm LAMOST} = (0.91 \pm 0.06) \cdot \log ([NII]6584/H_{\alpha})_{\rm SDSS} - (0.02 \pm 0.02)
\end{equation}

\section{Results and Discussions}
\subsection{Optical data}
We find a total of 20 double-peaked NEL objects, 15 galaxies and 5 QSO, listed in Table \ref{tab1-list}.
Among them, 10 have already been published by other authors, the remaining 10 being first discoveries. Three new spectra were first observed by LAMOST, see the Fig. \ref{fig3-spectra}. All our candidates have a high S/N with an average r-band magnitude (in SDSS system) of $\overline{r}$ = 17.17.

\begin{table}[htb!]
\caption[]{List of double-peaked NEL sample in LAMOST DR1}
\label{tab1-list}
\centering
\setlength{\tabcolsep}{2pt}
\begin{tabular}{cccccccccc}
    \hline
    \hline\noalign{\smallskip}
Designation & Obs.Date & MJD & Plate & SPID & FIB & Type & z & r & SDSS\\
\scriptsize(1) & \scriptsize(2) & \scriptsize(3) & \scriptsize(4) & \scriptsize(5) & \scriptsize(6) & \scriptsize(7) & \scriptsize(8) & \scriptsize(9)\\
\hline\noalign{\smallskip}
J005407.03+073611.9 & 20121108 & 56240 & EG010249N073002F & 14 & 047 & G & 0.10719 & 17.05 & 0\\
J023658.06+024217.9 & 20121114 (*)& 56246 & EG023131N032619F & 06 & 232 & G & 0.08744 & 16.17 & 1\\
J023832.67+023349.1 & 20121114 & 56246 & EG023131N032619F & 06 & 206 & G & 0.20780 & 17.13 & 0\\
J083425.28+283451.3 & 20130113 & 56306 & HD083217N291909F01 & 08 & 157 & G & 0.10253 & 17.04 & 1\\
J091646.03+283526.7 & 20120201 & 55959 & B5595902 & 15 & 006 & G & 0.14233 & 16.63 & 1\\
J094430.79+435421.4 & 20130215 & 56339 & HD094642N450651F01 & 05 & 088 & Q & 0.58779 & 17.77 & 1\\
J100708.01+242039.0 & 20130401 & 56384 & HD100153N235852F01 & 09 & 134 & Q & 0.54355 & 18.81 & 1\\
J104718.31+254348.3 & 20130210 & 56334 & HD104049N254200F01 & 09 & 237 & Q & 0.62712 & 18.72 & 1\\
J113630.61+135848.8 & 20130214 & 56338 & HD112941N152447F01 & 07 & 088 & G & 0.08169 & 16.91 & 1\\
J121342.90+422202.8	& 20130307 (**) & 56359 & HD121906N401846F01	& 16 & 245 & G & 0.07525 & 15.46 & 1\\
J123314.49+262624.9 & 20130111 & 56304 & HD122624N271605F01 & 06 & 006 & Q & 0.57025 & 18.58 & 1\\
J131434.73+563419.3 & 20130430 & 56413 & HD132545N565813F01 & 10 & 239 & G & 0.14407 & 17.39 & 1\\
J133730.29-002525.4 & 20130406 & 56389 & HD134427N004207F01 & 10 & 050 & G & 0.17212 & 17.11 & 1\\
J133737.82+555816.7 & 20130430 & 56413 & HD132545N565813F01 & 06 & 023 & Q & 0.47461 & 19.26 & 1\\
J135207.73+052555.8 & 20130409 & 56392 & HD135024N052949M01 & 04 & 239 & G & 0.07892 & 15.01 & 1\\
J135646.10+102609.0 & 20120514 & 56062 & VB3\_210N09\_V2 & 14 & 091 &  G & 0.12313 & 15.81 & 1\\
J140225.68+465817.4 & 20130310 & 56362 & HD141351N461930F01 & 14 & 218 & G & 0.12759 & 16.79 & 1\\
J140231.28+184807.8 & 20130210 & 56334 & HD140137N164527F01 & 11 & 118 & G & 0.13970 & 18.95 & 0\\
J150501.56+371311.7 & 20130402 & 56385 & HD145553N362548F01 & 13 & 109 & G & 0.06517 & 16.25 & 1\\
J232703.17+004256.7 & 20121024 & 56225 & EG232111N021150M01 & 07 & 073 & G & 0.06601 & 16.47 & 1\\
\noalign{\smallskip}

   \hline
\end{tabular}
\tablecomments{0.86\textwidth}{
Column 1: LAMOST DR1 designation hhmmss.ss+ddmmss.s (J2000.0); Column 2: Date of observation; Columns 3-6: the MJD, Plate, Spectrograph id and Fiber id in LAMOST Sky Survey; Column 7: Type: G: galaxy nucleus, Q: QSO;
Column 8: redshift; Column 9: apparent magnitude in r-band (in SDSS system); Column 10: 1 or 0 means that the object has a spectrum in SDSS DR9  or not.\\
(*) observed also on 20130501, (**) observed also on 20121113
}
\end{table}

\begin{figure}
\centering
\subfigure{
\begin{minipage}[b]{0.48\textwidth}
\includegraphics[width=1\textwidth]{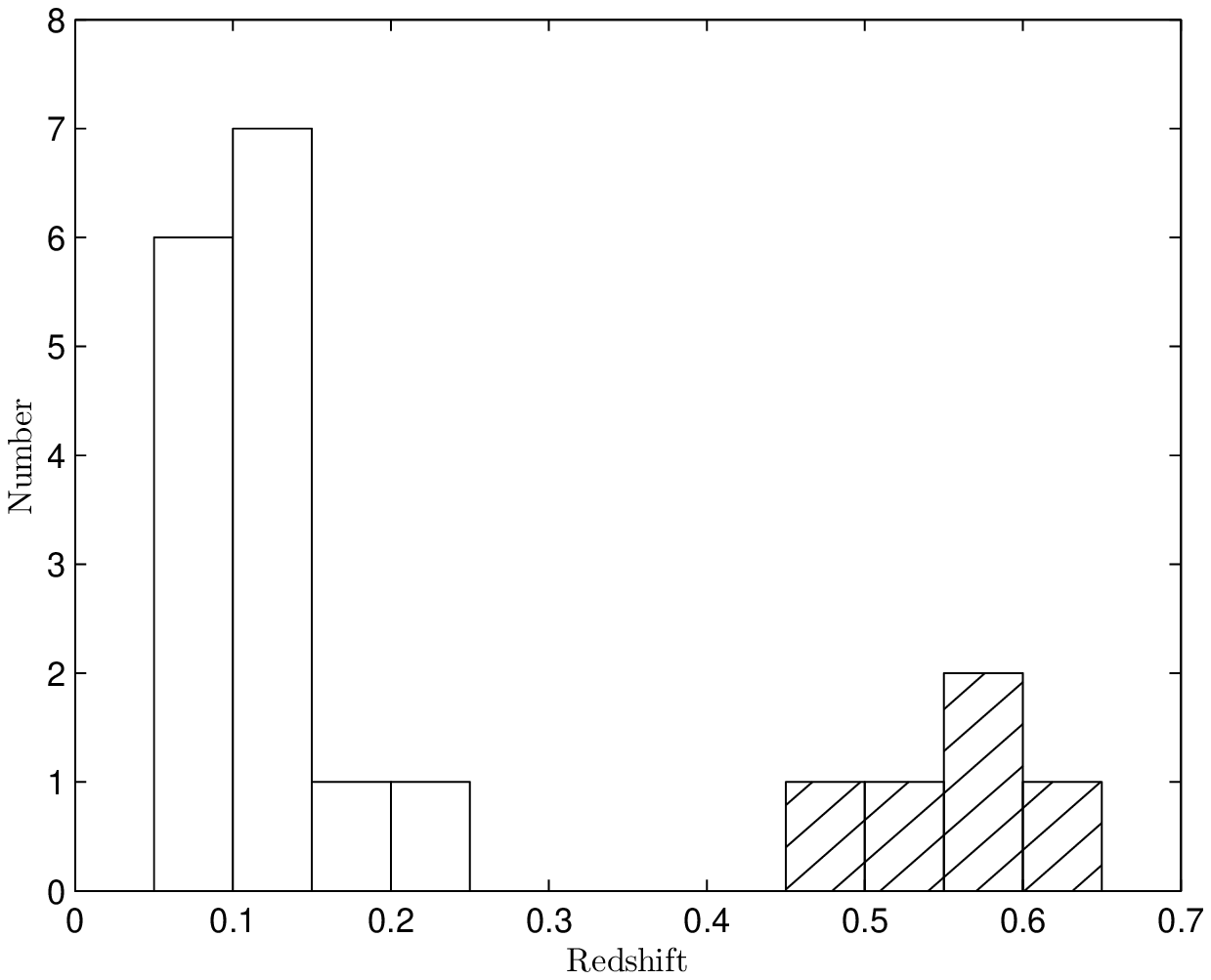} \\
\includegraphics[width=1\textwidth]{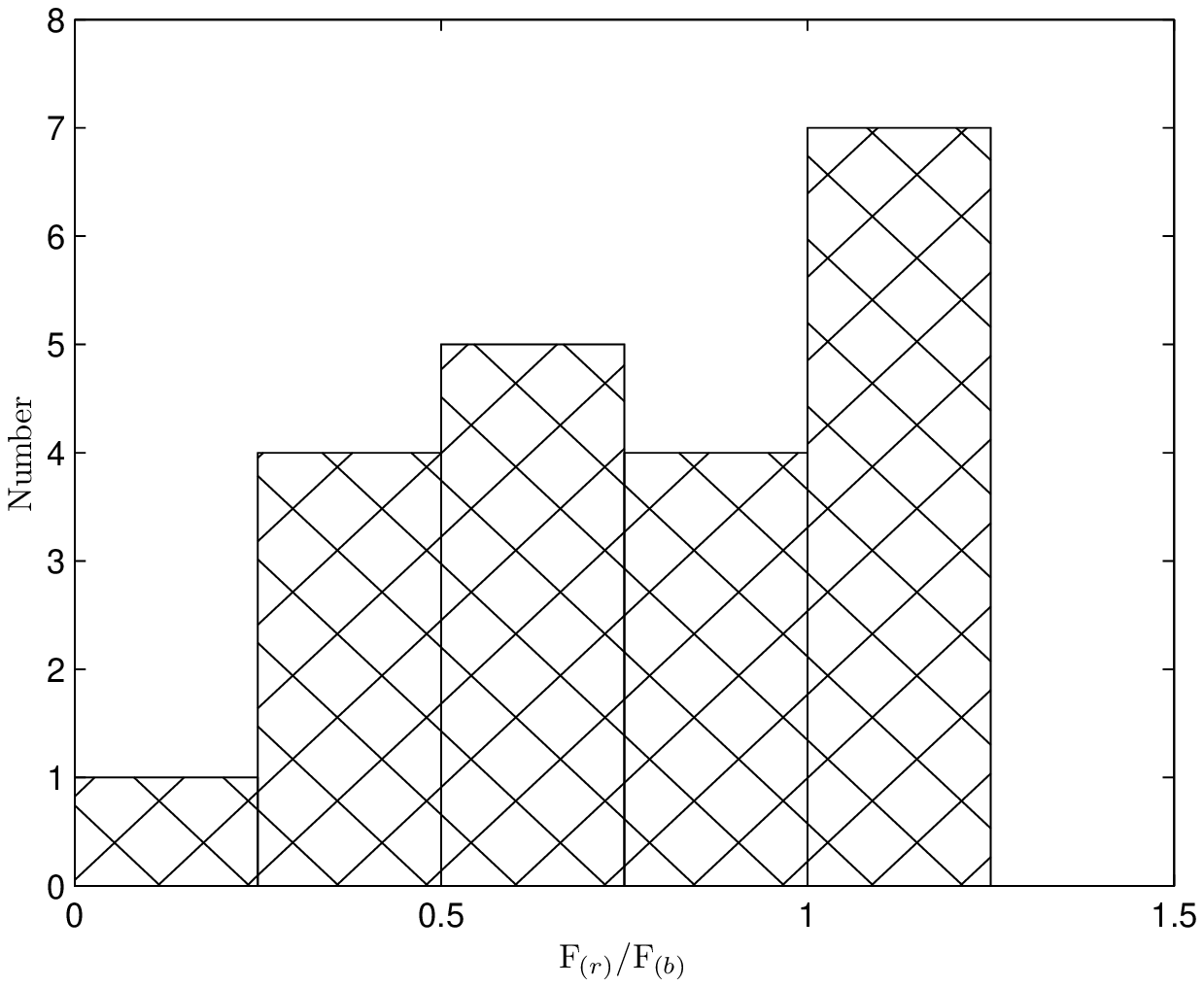}
\end{minipage}
}
\subfigure{
\begin{minipage}[b]{0.48\textwidth}
\includegraphics[width=1\textwidth]{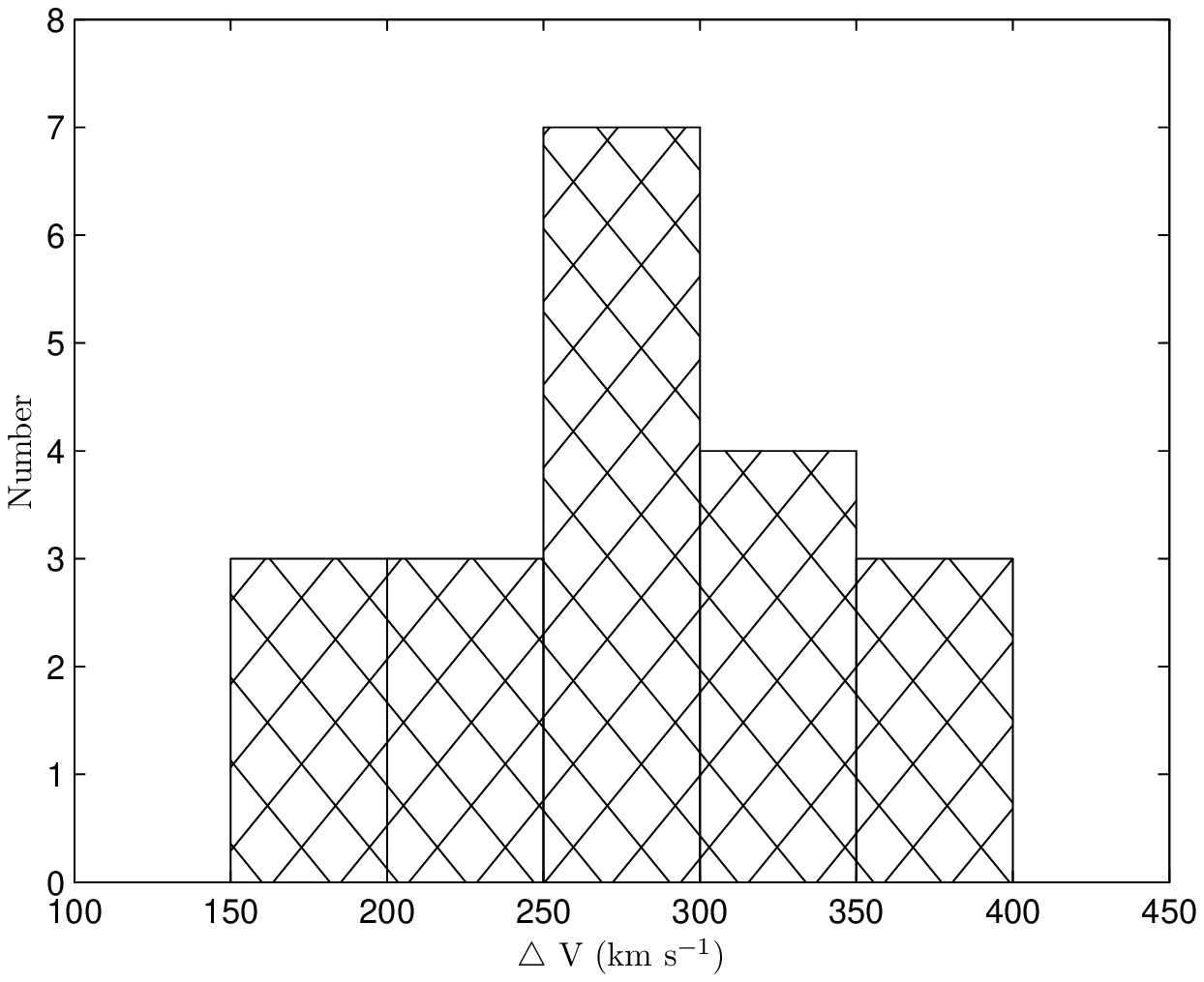} \\
\includegraphics[width=1\textwidth]{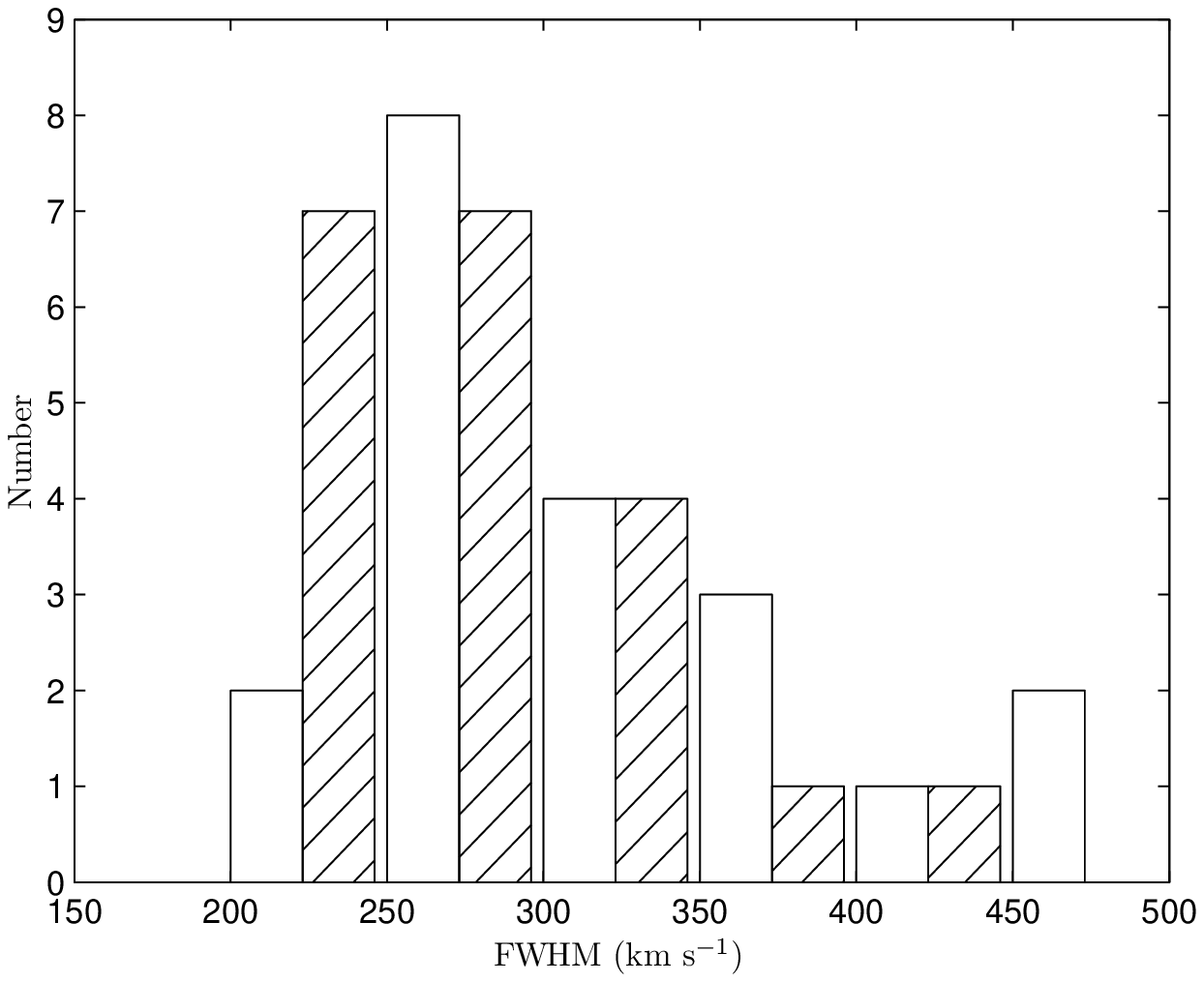}
\end{minipage}
}
\caption{The distribution of our double-peaked objects.
Top left (a): redshift distribution of galaxies(empty bars) and QSOs(line bars);
Top right (b): distribution of velocity difference between NEL components;
Bottom left (c): flux ratio of the red and blue \oiii\ components;
Bottom right (d): FWHM distribution of blue (empty bars) and red (line bars) components;}
	\label{fig4-dis}
\end{figure}

Fig. \ref{fig4-dis}a shows their redshift distribution, which is strongly bimodal because of our selection criteria. Objects with $0.5 \leq z \leq 0.7$ are all optically unresolved QSO, for which the emission complex \ha\ - \nii\ is shifted beyond the available spectral range. The double-peaked NEL character for these objects is only determined from the profiles of \oiii $\lambda\lambda\ 4959, 5007$ and the narrow component of \hb. An additional constraint is added to the initial trimming criteria (section 2.2.1 above) because of poor quality of LAMOST spectra beyond 8500 $\AA$, as a redshift cut at $z = 0.7$ so that \oiii $\lambda 5007$ is measurable in good conditions.
Objects with $z \leq 0.2$ are resolved galaxies on the SDSS images. To ensure a reliable identification of double NEL,
we require that not only \hb\ and \oiii\ $\lambda\lambda\ 4959, 5007$, but also \nii $\lambda\lambda\ 6548, 6584$, \ha\ and \sii $\lambda\lambda\ 6717, 6731$ all present double-peaked narrow-line profiles,
and for these we require as additional constraint that \sii $\lambda\ 6731$ $<$ 8500$\AA$. Note that this difference in selection criteria between galaxies and QSO is motivated by the data quality: in LAMOST DR1, the bulk of the redshift distribution of QSOs is clearly higher than that of the galaxies (at a given average magnitude) and the signal-to-noise of QSO spectra is generally better than that of galaxy nuclei, probably because of illumination conditions of the entrance face of the fiber.

Fig. \ref{fig4-dis}b shows the distribution of the velocity differences between the two components of the NEL. The distribution is roughly symmetric around a mean value of  $ <\Delta V>$ = 280 \kms . There is no significant difference in the observed $ <\Delta V>$ between QSO and galaxies. The spectral resolution of LAMOST corresponds to an average instrumental FWHM of 3.5  $A$, i.e. 200 \kms on the \oiii $\lambda 5007$ at the average redshift of the galaxy subsample (for higher z or lines in the red range the instrumental constraints are less stringent). The accuracy of centering of a good s/n gaussian component having such an FWHM, with usual computer algorithms, is roughly 1/10 of the FWHM, degrading when the s/n weakens. Hence we are confident that all our detected double peaked NEL listed in Table \ref{tab2-emission} are reliable. Further, the same procedures applied to the 17 objects that have a spectrum in the SDSS DR9 yielded consistent results with those listed in Table \ref{tab2-emission}, confirming the double peaked character of the NEL. Finally, two objects, J121342+4222 and J023658+0242 have been observed twice by LAMOST and the data as well as the fitting results of the two-epoch spectra are consistent.

Fig \ref{fig4-dis}c shows the distribution of the intensity ratios of the ''blue'' and ''red'' NEL velocity components in \oiii $\lambda 5007$ line.
About 50\% objects in our sample have a flux ratio between 0.75 and 1.25. Smith et al. (\cite{smith2012}) argue that double peaked AGN in which the two narrow line components have closely similar intensity often represent rotating disks, and are inconsistent with a black hole binary scenario.
There remain 50\% objects, in which the ''blue'' component flux is stronger than the ''red'' component one.
If a simple interaction between the a radio jet and the NLR was at work, (see also 3.4.2 below) the ''blue'' component should be systematically stronger than the ''red'' component.

\begin{figure}
   \centering
   \includegraphics[width=0.8\textwidth]{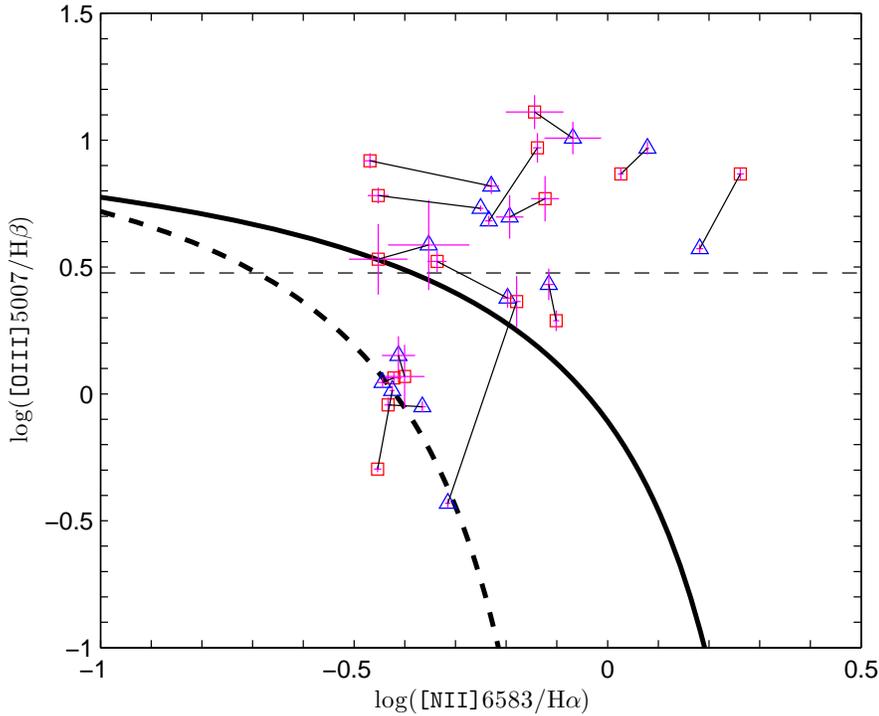}
   \caption{The BPT diagnostic diagram for objects with measured  \nii\ and \ha\ emission lines (14 objects).
   For each object, a blue triangle indicates the low velocity  component and the red square indicates the high velocity one.
   Two components belonging to the same source are connected by a thin continuous black line. The error bars are shown in magenta.
   The solid curve defined by Kewley et al. (\cite{ref56}) and the dashed curve defined by Kauffmann et al. (\cite{ref55})
   show the separation between star-forming galaxies, composite galaxies, and AGNs.
   The horizontal dotted line are defined by \oiii\ / \hb\ = 3. This line is often suggested to separate Seyferts from LINERS.}
   \label{fig5-bpt}
   \end{figure}

Fig \ref{fig4-dis}d shows the distributions of the FWHM of the ''blue'' and ''red'' NEL velocity components. Most objects exhibit intrinsically narrow lines, with FWHM $\leq$ 300 \kms on both components,

Fig. \ref{fig5-bpt} shows, for the objects belonging to the {\it galaxy} subsample, the location of the two NEL components in the Baldwin, Philips \& Terlevich (\cite{ref70}) (hereafter BPT) emission line diagnostic diagram. We recall that there is still some debate about the accuracy of the AGN versus starburst (or HII-like) spectrum distinction in the BPT diagram, the theoretical dividing line as derived by Kewley et al. (\cite{ref56}) being largely offset from the empirical dividing line derived by Kauffmann et al.(\cite{ref55}) from a very large sample of emission-line galaxies.
In 7 objects, both components are clearly located in the Kewley et al. AGN region, implying that the ionization of their narrow-line plasma are dominated by non-thermal sources. Note that, from this diagram only, radiative shock contribution cannot be evaluated quantitatively. 3 other objects have their two components close to the Kewley et al separation line between AGN-like spectra and thermal HII-like spectra, but inside the Kauffmann et al. AGN region. For these objects, a significant contribution of thermal photoionization by ongoing massive star formation is possible. One object (J083425+283451) exhibits widely different behaviour in its two NEL components, the ''blue'' one being probably largely dominated by thermal photoionization, the ''red'' one being transitional, or AGN-like, on the Kewley et al. separation line. The 4 remaining objects have their NEL components clustered in a small area of the diagram, with a small dispersion around \oiii / \hb   = 1 for \nii / \ha = 0.4  , i.e. close to the Kauffmann et al. AGN / HII separation line. These objects are likely to be largely dominated by massive star formation, non-thermal AGN-like ionization playing only a marginal role, if any, but weak shocks being still possible. For all objects and all components, the ionization parameter is quite strong. These results are summarized in Table \ref{tab2-emission}, column 10.

\begin{landscape}
\begin{table}[htb!]
\caption{Emission lines properties of the double-peaked emission line sample}
\label{tab2-emission}
\centering
\setlength{\tabcolsep}{2pt}
\renewcommand{\arraystretch}{1.5}
\begin{tabular}{@{}ccccccccccc@{}}

   \hline\noalign{\smallskip}
Designation & NEL model & $\bigtriangleup V$ & $FWHM_{b}$ & $FWHM_{r}$ & $F{^b_{\texttt{\oiii}}}$/$F{^b_{H_{\beta}}}$ & $F{^r_{\texttt{\oiii}}}$/$F{^r_{H_{\beta}}}$ & $F{^b_{\texttt{\nii}}}$/$F{^b_{H_{\alpha}}}$ & $F{^r_{\texttt{\nii}}}$/$F{^r_{H_{\alpha}}}$ & Type & Ref\\
\scriptsize(1) & \scriptsize(2) & \scriptsize(3) & \scriptsize(4) & \scriptsize(5) & \scriptsize(6) & \scriptsize(7) & \scriptsize(8) & \scriptsize(9) &
\scriptsize(10) & \scriptsize(11)\\
      \hline\noalign{\smallskip}

J005407.03+073611.9 & 2G &$ 279 \pm 44 $&$ 301 \pm 43 $&$ 224 \pm 48 $&$ 1.42 \pm 0.25 $&$ 1.17 \pm 0.34 $&$ 0.39 \pm 0.03 $&$ 0.40 \pm 0.04 $& SF + SF & - \\
J023658.06+024217.9 & 2G + W &$ 250 \pm 49 $&$ 236 \pm 44 $&$ 241 \pm 33 $&$ 4.99 \pm 0.99 $&$ 5.88 \pm 1.21 $&$ 0.64 \pm 0.04 $&$ 0.75 \pm 0.05 $& AGN + AGN & - \\
J023832.67+023349.1 &  2G &$ 378 \pm 83 $&$ 397 \pm 53 $&$ 325 \pm 39 $&$ 10.20 \pm 1.49 $&$ 12.91 \pm 1.99 $&$ 0.85 \pm 0.11 $&$ 0.72 \pm 0.09 $& AGN + AGN & - \\
J083425.28+283451.3 &   2G  &$ 294 \pm 43 $&$ 296 \pm 16 $&$ 346 \pm 53 $&$ 0.37 \pm 0.02 $&$ 2.31 \pm 0.54 $&$ 0.48 \pm 0.01 $&$ 0.66 \pm 0.01 $& SF + Comp & (4b) \\
J091646.03+283526.7 &  2G + W &$ 392 \pm 17 $&$ 391 \pm 15 $&$ 251 \pm 29 $&$ 5.39 \pm 0.14 $&$ 6.05 \pm 0.44 $&$ 0.56 \pm 0.01 $&$ 0.35 \pm 0.02 $& AGN + AGN & (2) \\
J094430.79+435421.4 &   2G + W &$ 306 \pm 16 $&$ 280 \pm 15 $&$ 255 \pm 21 $&$ 4.81 \pm 1.03 $&$ 9.24 \pm 4.32 $&$ - $&$ - $& - & - \\
J100708.01+242039.0 &  2G + W  &$ 340 \pm 25 $&$ 373 \pm 80 $&$ 230 \pm 115 $&$ 3.87 \pm 1.74 $&$ 3.43 \pm 0.71 $&$ - $&$ - $& - & (3) \\
J104718.31+254348.3 &   2G + W  &$ 249 \pm 58 $&$ 252 \pm 46 $&$ 270 \pm 59 $&$ 6.31 \pm 2.27 $&$ 13.38 \pm 9.54 $&$ - $&$ - $& - & - \\
J113630.61+135848.8 &  2G  &$ 192 \pm 39 $&$ 322 \pm 33 $&$ 301 \pm 30 $&$ 2.70 \pm 0.39 $&$ 1.94 \pm 0.18 $&$ 0.77 \pm 0.02 $&$ 0.79 \pm 0.01 $& Comp + Comp & (1) \\
J121342.90+422202.8 &  2G  &$ 349 \pm 88 $&$ 414 \pm 72 $&$ 425 \pm 44 $&$ 9.30 \pm 0.54 $&$ 7.36 \pm 0.43 $&$ 1.20 \pm 0.01 $&$ 1.06 \pm 0.02 $& AGN + AGN & - \\
J123314.49+262624.9 &  2G + W &$ 253 \pm 16 $&$ 297 \pm 23 $&$ 239 \pm 17 $&$ 7.56 \pm 0.66 $&$ 7.41 \pm 0.63 $&$ - $&$ - $& - & - \\
J131434.73+563419.3 &  2G  &$ 181 \pm 21 $&$ 282 \pm 25 $&$ 251 \pm 19 $&$ 0.89 \pm 0.05 $&$ 0.91 \pm 0.06 $&$ 0.43 \pm 0.01 $&$ 0.37 \pm 0.01 $& SF + SF & (4b) \\
J133730.29-002525.4 &  2G  &$ 239 \pm 18 $&$ 267 \pm 19 $&$ 230 \pm 18 $&$ 2.39 \pm 0.21 $&$ 3.33 \pm 0.31 $&$ 0.64 \pm 0.02 $&$ 0.46 \pm 0.02 $& Comp + Comp & (4b) \\
J133737.82+555816.7 &  2G + W  &$ 265 \pm 28 $&$ 314 \pm 21 $&$ 327 \pm 31 $&$ 8.92 \pm 4.83 $&$ 3.05 \pm 0.65 $&$ - $&$ - $& - & - \\
J135207.73+052555.8 &  2G  &$ 331 \pm 16 $&$ 450 \pm 20 $&$ 263 \pm 9 $&$ 3.74 \pm 0.15 $&$ 7.36 \pm 0.36 $&$ 1.52 \pm 0.03 $&$ 1.83 \pm 0.03 $& AGN + AGN & (1), (4a) \\
J135646.10+102609.0 &  2G + W &$ 386 \pm 23 $&$ 484 \pm 23 $&$ 394 \pm 26 $&$ 6.59 \pm 0.46 $&$ 8.29 \pm 0.57 $&$ 0.59 \pm 0.02* $&$ 0.34 \pm 0.01* $& AGN + AGN & (2) \\
J140225.68+465817.4 &   2G  &$ 224 \pm 42 $&$ 281 \pm 20 $&$ 260 \pm 24 $&$ 1.03 \pm 0.04 $&$ 0.51 \pm 0.02 $&$ 0.38 \pm 0.01 $&$ 0.35 \pm 0.01 $& SF + SF & - \\
J140231.28+184807.8 &  2G + W & $ 156 \pm 109 $&$ 203 \pm 74 $&$ 207 \pm 72 $&$ 3.86 \pm 1.58 $&$ 3.39 \pm 1.09 $&$ 0.44 \pm 0.08 $&$ 0.35 \pm 0.05 $& Comp + Comp & - \\
J150501.56+371311.7 &  2G  &$ 256 \pm 39 $&$ 287 \pm 19 $&$ 292 \pm 37 $&$ 4.81 \pm 0.23 $&$ 9.33 \pm 1.25 $&$ 0.58 \pm 0.01 $&$ 0.73 \pm 0.02 $& AGN + AGN & (4b) \\
J232703.17+004256.7 &  2G   &$ 267 \pm 18 $&$ 332 \pm 16 $&$ 235 \pm 16 $&$ 1.11 \pm 0.05 $&$ 1.16 \pm 0.06 $&$ 0.36 \pm 0.01 $&$ 0.38 \pm 0.01 $& SF + SF & (4a) \\

\hline\hline
\end{tabular}
\tablecomments{0.86\textwidth}{
Column 1: LAMOST DR1 designation hhmmss.ss+ddmmss.s (J2000.0)\\
Column 2: best NEL model: 2G: 2 gaussians, 2G+W :  2 gaussian + a broader gaussian ''wing''\\
Column 3: velocity difference in \kms  between the two NEL components, in the rest frame defined by the redshift listed in Table \ref{tab1-list} \\
Columns 4-5: FWHMs of blue and red components, in units of \kms \\
Column 6-7: flux ratio of \oiii\ $\lambda\ 5007$ and \hb \\
Column 8-9: flux ratio of \nii\ $\lambda\ 6583$ and \ha. Note * indicate that the values are from the spectrum of SDSS for technical reason.  \\
Column 10: NEL components location in BPT diagram: AGN: AGN-like line ratios, SF: likely dominated by massive star formation, Comp: on the Kewley et al. (\cite{ref56}) AGN-HII separation line, possibly composite or transitional (in the Kauffmann et al. (\cite{ref55}) these would be classified AGN) \\
Column 11: (1): Wang et al. \cite{ref5}, (2): Liu et al. \cite{ref65}, (3): Smith et al. \cite{ref8}, (4) Ge et al. \cite{ref9}: a) listed as double-peaked NEL, b) listed as asymmetric NEL
}

\end{table}
\end{landscape}

\begin{table}[htb!]
\caption[]{Additional FIRST radio data for our double-peaked NEL sample from LAMOST DR1}
\label{tab3-first}
\centering

\begin{tabular}{cccccc}
    \hline
    \hline\noalign{\smallskip}
Designation & Max. flux density & Integrated flux & s/n & Note & Ref. \\
\scriptsize(1) & \scriptsize(2) & \scriptsize(3) & \scriptsize(4) & \scriptsize(5) & \scriptsize(6)\\
\hline\noalign{\smallskip}
J005407.03+073611.9 & 0.46 & 0.48  & 4.0 &  P  & (2) \\
J023658.06+024217.9 & 0.36 & 0.38  & 3.0 &  P  & (2) \\
J023832.67+023349.1 & 23.2 & 24.94 & 142 & Ext. & (1),(3) \\
J091646.03+283526.7 & 5.25 & 5.76  & 35  &  P  &  (1),(3) \\
J094430.79+435421.4 & 0.61 & 0.94  & 5.0 & Ext?, (*) & (2) \\
J113630.61+135848.8 & 0.42 & 0.65  & 3.5 &  P & (2) \\
J121342.90+422202.8	& 1.84 & 1.58  & 12	 &  P & (1) \\
J131434.73+563419.3 & 0.74 & 0.99  & 5.2 & P & (2) \\
J135207.73+052555.8 & 0.64 & 1.26  & 4.5  & P & (2) \\
J135646.10+102609.0 & 56.5 & 59.58 & 362 & P & (1),(3) \\
J140225.68+465817.4 & 0.64 & 0.52  & 4.5  & P & (2)  \\
J140231.28+184807.8 & 0.90 & 1.04  & 6.2  & P & (1) \\
\noalign{\smallskip}
   \hline
\end{tabular}
\tablecomments{0.86\textwidth}{
Column 1: LAMOST DR1 designation hhmmss.ss+ddmmss.s (J2000.0); Column 2: Maximum flux density on source, in $mJy/beam$; Columns 3: integrated flux in $mJy$ from FIRST catalog; Column 4: local signal-to-noise on source;
Column 5: P: point-source, Ext: resolved extended source; Column 6: reference: (1): FIRST source catalog, (2): present work, (3) also in NVSS \\
(*) indicates possible sidelobe contamination on FIRST map.
}

\end{table}

\subsection{Additional radio data at 1.4 $GHz$ and WISE infrared data}

We have searched the FIRST radio survey for counterparts at 1.4 $GHz$
of our candidates. 5 sources have been found in the FIRST source catalog, whose flux limit is on average 1 $mJy$ across the surveyed area. However, from a close examination of the FIRST image cutouts, 7 more detections were found, all with close positional coincidences (less than 2 ''). 2 of these remain quite marginal and the derived fluxes may have large uncertainties. All sources are unresolved except J023832+023349 (alias PKS 0235+023) which is a radiogalaxy exhibiting structure in its image. J094430+435421 has a slightly NS elongated radio image but appears located in a quite noisy area polluted by ripples that could be due to sidelobes effects from a distant bright source in the same field.
The radio data derived from FIRST are listed in Table \ref{tab3-first}.
We have also examined the NVSS survey, but the angular resolution is much better in FIRST (5 '' instead of 45 ''), although the sensitivity of NVSS to weak extended structures may be better. In the NVSS catalog, 3 sources are present, namely J023832+023349, J091646+283526 and J135646+102609.\\
\indent The WISE space experiment (Wright et al. \cite{ref67}, Mainzer et al. \cite{ref68}) has provided an all-sky map and source catalog in
4 spectral bands centered at 3.4, 4.6, 12 and 22 microns. The two first bands are usually dominated by starlight, the two longest by hot interstellar dust.
The spatial resolution is excellent ( ~6" in the three shortest bands, 12" in the fourth one) and the sensitivity very high.
We have searched the AllWISE catalog for counterparts to our objects. The results are given in Table \ref{tab4-wise},
where WISE magnitudes have been converted to fluxes in millijanskys using the recommendations of Jarrett et al. (\cite{ref69}). All sources are detected in the first three bands. In the 22 $\mu$ band 4 objects  have a s/n below the formal detection threshold.
3 sources (J023658.06+024217.9, J121342.90+422202.8, J135207.73+052555.8) are fitted with two gaussian point-source components by the catalog construction pipeline, implying resolved structure.

\begin{table}[htb!]
\caption[]{Additional WISE infrared data for our double-peaked NEL sample from LAMOST DR1}
\label{tab4-wise}
\centering

\begin{tabular}{ccccc}
    \hline
    \hline\noalign{\smallskip}
Designation & 3.4 $\mu$ flux & 4.6 $\mu$ flux & 12 $\mu$ flux & 22 $\mu$ flux \\
\scriptsize(1) & \scriptsize(2) & \scriptsize(3) & \scriptsize(4) & \scriptsize(5) \\
\hline\noalign{\smallskip}
J005407.03+073611.9 & $ 0.79 \pm 0.02 $ & $ 0.60 \pm 0.02 $ & $ 3.22 \pm 0.18 $ & $ 8.00 \pm 1.21 $ \\
J023658.06+024217.9 &  $ 2.20 \pm 0.07 $ & $ 1.50 \pm 0.06 $ & $ 7.55 \pm 0.35 $ & $ 9.70 \pm 1.80 $  \\
J023832.67+023349.1 &  $ 4.00 \pm 0.09 $ & $ 5.49 \pm 0.12 $ & $ 11.84 \pm 0.25 $ & $ 34.15 \pm 1.40 $  \\
J083425.28+283451.3 &  $ 0.80 \pm 0.02 $ & $ 0.57 \pm 0.02 $ & $ 2.45 \pm 0.16 $ & undetected \\
J091646.03+283526.7 &  $ 5.17 \pm 0.11 $ & $ 7.75 \pm 0.15 $ & $ 23.06 \pm 0.41 $ & $ 68.62 \pm 2.37 $  \\
J094430.79+435421.4 &  $ 0.97 \pm 0.02 $ & $ 1.35 \pm 0.03 $ & $ 2.22 \pm 0.15 $ & $ 6.85 \pm 1.16 $  \\
J100708.01+242039.0 & $ 0.69 \pm 0.02 $ & $ 1.00 \pm 0.03 $ & $ 1.88 \pm 0.15 $ & $ 4.25 \pm 1.37 $  \\
J104718.31+254348.3 &  $ 0.54 \pm 0.02 $ & $ 0.88 \pm 0.02 $ & $ 1.63 \pm 0.18 $ & $ 6.52 \pm 1.18 $ \\
J113630.61+135848.8 & $ 1.21 \pm 0.03 $ & $ 0.94 \pm 0.03 $ & $ 3.76 \pm 0.20 $ & $ 5.92 \pm 1.51 $ \\
J121342.90+422202.8	& $ 2.75 \pm 0.06 $ & $ 1.97 \pm 0.05 $ & $ 6.50 \pm 0.25 $ & $ 19.65 \pm 1.33 $ \\
J123314.49+262624.9 & $ 0.40 \pm 0.01 $ & $ 0.56 \pm 0.02 $ & $ 1.25 \pm 0.14 $ & undetected \\
J131434.73+563419.3 & $ 0.71 \pm 0.02 $ & $ 0.66 \pm 0.02 $ & $ 5.39 \pm 0.16 $ & $ 9.68 \pm 0.93 $ \\
J133730.29-002525.4 & $ 0.68 \pm 0.02 $ & $ 0.58 \pm 0.02 $ & $ 3.37 \pm 0.16 $ & $ 5.91 \pm 0.86 $ \\
J133737.82+555816.7 & $ 0.22 \pm 0.01 $ & $ 0.28 \pm 0.01 $ & $ 0.93 \pm 0.12 $ & undetected \\
J135207.73+052555.8 & $ 5.25 \pm 0.15 $ & $ 2.53 \pm 0.10 $ & $ 6.63 \pm 0.30 $ & $ 9.50 \pm 1.80 $ \\
J135646.10+102609.0 & $ 2.23 \pm 0.05 $ & $ 4.47 \pm 0.09 $ & $ 29.19 \pm 0.46 $ & $ 178.83 \pm 4.37 $ \\
J140225.68+465817.4 & $ 0.86 \pm 0.02 $ & $ 0.66 \pm 0.02 $ & $ 5.18 \pm 0.16 $ & $ 8.42 \pm 0.77 $  \\
J140231.28+184807.8 & $ 0.36 \pm 0.01 $ & $ 0.48 \pm 0.02 $ & $ 2.45 \pm 0.12 $ & $ 9.16 \pm 0.82 $ \\
J150501.56+371311.7 & $ 3.93 \pm 0.09 $ & $ 4.67 \pm 0.09 $ & $ 11.42 \pm 0.25 $ & $ 27.30 \pm 1.04 $ \\
J232703.17+004256.7 & $ 1.06 \pm 0.03 $ & $ 0.68 \pm 0.02 $ & $ 1.77 \pm 0.17 $ & undetected \\

\noalign{\smallskip}
   \hline
\end{tabular}
\tablecomments{0.86\textwidth}{
Column 1: LAMOST DR1 designation hhmmss.ss+ddmmss.s (J2000.0); Column 2-5: WISE flux of source, in $mJy$, computed from the magnitudes in AllWISE Catalog, for passbands 3.4, 4.6, 12 and 22 $\mu$. \\
For three sources resolved in subcomponents, the fluxes of components were added. The 22 $\mu$ fluxes have been corrected following Jarrett et al. (\cite{ref69}). A source is considered as undetected if its s/n is lower than 3 \\
}
\end{table}

\subsection{Notes on individual objects}
We report here some remarks from SDSS images and references on previous work related to two objects.

J005407+073611: no morphological information available (technical problems in SDSS image)

J023658+024217: an Sb spiral with a bright nucleus. In close group with several galaxies whose appearance suggest similar redshift. The absorption lines could also be double, but this needs confirmation with higher s/n. The SDSS redshift corresponds to the velocity of the blue NEL component.

J023832+023349, (PKS 0235+023), is a broad line Seyfert 1 galaxy,
probably of SBa type, whose  \ha\ broad line exhibits double peaks or twin shoulders. This was fitted quite well, on 1991 observations, with a model attributing the broad emission to a circular, relativistic, Keplerian disk  (Eracleous \& Halpern \cite{ref37, ref49}).
However, Gezari et al. (\cite{ref50}) presented a long-term monitoring of the double-peaked broad \ha\ emission lines
and found a dramatic change in profile shape after 1991 leading to
consider the simple circular disk model as insufficient for interpretation of the spectra taken after 1991. The present LAMOST spectrum not only shows a double peak on \ha\ broad component, but also double peaks on all $narrow$ lines including [OII], [NeIII] and [NeV] as well as the Balmer series. The broad component seems highly obscured by internal extinction, since it is almost invisible in \hb. It is clear that this object deserves detailed follow-up observations.

J083425+283451: a SBb spiral with a low-excitation emission spectrum.

J091646+283526 is a quite amorphous galaxy with a perturbed morphology. The bright nucleus is close (2-3'') to a second, much redder one. A moderately broad component is present in the Balmer emission lines. The excitation is very strong with bright HeII and [NeV] lines (Seyfert 1.5). This object is a very good merger candidate. The NVSS map shows an extended source with two components, one being $\pm$ coincident with the optical object.
Fu et al. (\cite{ref22}) place J091646+283526 in the category of unresolved narrow-line regions from integral-field spectroscopy.
They argue that most of these spatially unresolved double-peaked NEL AGNs are aligned or young outflows.

J094430+435421: QSO whose NELs are well splitted. The \hb\ line is  coincident with the telluric A band of oxygen making spectrophotometry uncertain.

J100708+242039 and J104718+254348 are QSOs on which few information is available. The latter is an X-ray source.

J113630+135848: a $\sim$ amorphous galaxy with possible pair of opposite, low surface brightness, blue diffuse extensions.

J121342+422202: a galaxy with a distended asymmetric envelope. The \ha\ range is lacking on the SDSS spectrum. A small diffuse object is very close. The galaxy is member of a group of more than 10 objects of similar redshift, among which a Seyfert 1 at 4.1 $arcmin$ N. Possible merger or post-merger.

J123314+262624: QSO, no information.

J131434+563419: peculiar asymmetric-shaped, possibly barred, galaxy with a blue extension that could be a tidal tail. Possible merger candidate. In a loose group with other, similar redshift, galaxies.

J133730-002525: a spiralgalaxy (Sbc/Sc) seen close to face-on, with many faint objects close-by (but no redshift information on them, except one at 5 $arcmin$ NE)

J133737+555816: QSO, no information

J135207+052555: typical luminous Seyfert 2 galaxy, of early-type (SO or SO/a). The nucleus appears unique at the spatial resolution of SDSS image. The double-lined character of the NEL is especially obvious on the \sii\ lines which appear as a triple peak. This object is a member of the Abell 1809 galaxy cluster.

J135646+102609 has aroused great interest in the recent literature.
Liu et al. (\cite{ref6}) included this object in a sample of type 2 AGNs with double-peaked NEL.
Its morphology is clearly disturbed, suggesting an ongoing merger.
In a loose cluster with several other less luminous objects. The red NEL component has the same redshift as the absorption lines of the stellar population. The presence of
two merging galactic nuclei has been demonstrated from optical (Greene et al. \cite{ref51, ref52}; Fu et al. \cite{ref22}) and  near-infrared (Shen et al. \cite{ref24}) observations.
Greene et al. (\cite{ref53}) have published two long-slit Magellan LDSS-3 spectra and additional spectral band data.
They consider that this object is a pair of interacting galaxies that hosts a luminous obscured quasar in its
northern nucleus. Note that the optical spectrum from LAMOST or SDSS is that of a typical
unobscured Seyfert 2 galaxy, all lines being narrow.

J140225+465817: a galaxy with perturbed morphology, with an extension (or projected close companion) along NE and irregular fuzzy extension along S. Possible merger candidate. Isolated.

J140231+184807: small apparent diameter galaxy, seems isolated.

J150501+371311: bright nucleus galaxy with a pear-shaped disk. A diffuse blue object at 25 $arcsec$ E, without redshift information.

J232703+004256: an $\sim$ amorphous galaxy (maybe Sa), forming an interactive pair with a companion at 25 $arcsec$ NE, of similar size and brightness. A diffuse matter bridge is seen between both objects, and a tidal tail appears towards SW. Single nucleus at SDSS resolution.

\subsection{Possible origins for double NEL structures}
\subsubsection{Black hole binaries}
One initial aim of this investigation was the quest for QSO and AGN candidates hosting binary black holes from the identification of double-peaked NEL in  optical spectra, as practised by Wang et al. (\cite{ref5}), Liu et al. (\cite{ref6,ref7}) and Smith et al. (\cite{ref8}). In our sample appeared galaxy nuclei candidates with consistent ''blue'' and ''red'' velocity systems in all the emission lines (i. e., \hb, \oiii\ $\lambda\lambda\ 4959, 5007$, \nii $\lambda\lambda\ 6548, 6584$, \ha\ and \sii $\lambda\lambda\ 6717, 6731$ ) of which five examples are shown in Fig 2. Among the 15 galaxies identified, 6 exhibit pairs of NEL components that may be unambiguously classified as AGN. The question remains if these double NEL systems correspond to double supermassive black hole engines. Besides, 5 QSO are also identified, among which one (J094430+435421) has spectacular double NEL components and has only a faint radio loudness.

 On a general ground, statistical analysis of double-peaked \oiii\ AGN samples by Shen et al. (\cite{ref24}) and Fu et al. (\cite{ref26}) tend to support the conclusion that a large fraction of the double-peak systems are driven by the NLR dynamics instead of the existence of black hole binaries. Ge et al. (\cite{ref9}) found only 54 dual-core galaxies with projected separations closer than 3" among a sample of 15600 objects with double-peak or asymmetric NEL, but the angular resolution of SDSS images is not appropriate to disentangle very close components.

 We also stress that previous studies of double-peaked NEL have focused on very obvious, well separated double lines in the line profiles, which have been observed since long but remain largely unexplained.
 In the present limited sample, with the availablable low or moderate spatial and spectral resolutions (optical spectra, SDSS images and FIRST radio maps),
 we do not find evidence for binary nuclei as exhibited by the already well-known binary BH candidates (Ge et al. \cite{ref9}),
 except in J091646+2835, which has two central condensations and has a double AGN-like spectrum NEL component.
 (Note that the well-studied J135646+102609 does not show up clearly as a binary nucleus merger on the SDSS images).
 This study hence must be completed with additional higher resolution including good seeing imagery and other multi-band data from follow-up observations.

\subsubsection{Interaction between jet and NLR}
In radio-loud AGNs, double narrow line structure may readily arise from the radio-jet interaction with the NLR clouds.
This operates mainly by ram pressure acceleration of the ionized gas clouds of the NLR, and can produce complicated NLR kinematics,
and especially double lines, due to projection on the line of sight of cone structure in the matter submitted to the interaction.
If the jet (whose direction points $\pm$ closely towards the observer) drags and accelerates NLR matter with it,
a Doppler blueshifted (with respect to the systemic redshift) ionized gas component may naturally appear in the spectrum.
Note that Smith et al. (\cite{ref8}) exclude the radio-loud AGNs from their candidates binary black hole
because they think that these objects are more or less dominated by the jet / NLR interaction.
The literature on these phenomena is quite extensive, and there have been extremely detailed studies on some nearby cases, most of them are Seyfert 2 galaxies
(See e.g. Whittle et al. \cite{ref42}, Rosario et al. \cite{ref43} with many references inside), for which high spatial resolution optical and radio mapping is possible.

In our double peaked NEL sample, the objects detected at 1.4 $GHz$ that have the largest radio flux (corresponding more or less to the largest radio luminosity) have their NEL components of type AGN + AGN with FWHM values larger than the average (typically larger than 350 \kms ). It should be useful to explore if this trend still holds on larger samples.

\subsubsection{Interaction with companion galaxies and merger candidates}
Galaxy collisions and subsequent merging are natural sources of double emission lines phenomena, depending on the spatial integration of the spectrograph entrance aperture, on the angle of projection of the interacting system on the line of sight and on the collision phase observed. Since interactions between gas-rich systems are likely to enhance the star formation, ionized gas could be observed in the two components of a strongly interacting system with velocity separations up to a few hundreds \kms. These interactions are accompanied by many morphological perturbations, such as tidal tails, matter bridges, deformation of disks, envelope asymmetric extensions, etc... Also, several authors have proposed that nuclear activity may sometimes be triggered by galaxy interactions.

In the present sample, examination of the SDSS images lead to find one clear interacting pair (J232703+0042), 1 merger (J135646+1026), 4 candidates mergers or possible candidates mergers (J091646+2835, J121342+4222, J131434+5634, J140225+4658), 2 more objects have perturbed morphology (J113630+1358, J150501+3713) and J135207+0525 belongs to a populated cluster in which encounters should have higher probability than in the field. Thus, 2/3 of our $galaxy$ sample are objects for which interaction is a plausible origin for NEL double structure. Verification of this hypothesis, its extension to other double-peaked NEL galaxy samples and better modeling cannot be done without follow-up observations, including deep imagery.

Regarding the QSO, for which morphology information is not available,
several previous authors (Hutchings et al. \cite{ref44}; Malkan, \cite{ref45}; Green and Yee \cite{ref46}) have noted that low-redshift quasars very often have faint companion galaxies with small projected angular separations. A scenario involving a collision between a giant AGN host galaxy and a gas-rich star-forming dwarf may lead to double structure in the NEL of the AGN: in the process of interaction, the companion is disrupted by the tidal field of the giant galaxy. Clumps of its interstellar matter may approach the AGN sufficiently close such that these clumps may be submitted to violent differential accelerations able to increase their velocity dispersion up to significantly higher values than the usual average quiet disk value. If interaction with a neighbor were the origin of the double structure in an AGN NEL, observations could disentangle the ionized gas of the companion, expected to have line ratios consistent with a normal massive stars photoionization, from the NLR component of the AGN in which the ionization is basically due to non-thermal source spectra. The present sample does not enable this for the QSO, because \ha\ is not observed.

\section{Summary}
We present a search for double-peaked NEL galaxies and AGNs in the LAMOST survey spectroscopic data base. Applying our method to LAMOST DR1, we find 20 candidates, among which 10 are first discoveries.
Spectra from SDSS, available for 17 objects, confirm their double-peaked narrow-line profiles. We report new weak radio continuum fluxes  and give the position of each NEL component in BPT diagnostic diagram for 15 objects.
We briefly discuss some possibilities about the origin of double components in emission lines, and underline that our galaxy subsample appears dominated by objects likely to be submitted to gravitational interaction with neighbours or ongoing merging processes.
The double-peaked sample can be used to study the dynamical processes of merging galaxies, jet-cloud interactions, and gas kinematics in the narrow-line regions.
With the ongoing progress of the LAMOST survey, we shall be able to enlarge the sample and perform statistical analysis on the population of double-peaked NEL galaxies and AGNs.
For the current sample, additional high resolution optical and multi-band follow-up observations are needed to understand the origins of
double components in emission lines.

\begin{acknowledgements}
\small{
This study is supported by the National Natural Science Foundation of China under
grant Nos. 10973021, 11233004, 61202315 and National Key Basic Research of China 2014CB845700.
Guoshoujing Telescope (the Large Sky Area Multi-Object Fiber Spectroscopic Telescope LAMOST)
is a National Major Scientific Project built by the Chinese Academy of Sciences.
Funding for the project has been provided by the National Development and Reform Commission.
LAMOST is operated and managed by the National Astronomical Observatories, Chinese Academy of Sciences.
The LAMOST Data Release 1 web site is \url{ http://data.lamost.org/dr1/}.
This publication makes use of data products from the Wide-field Infrared Survey Explorer,
which is a joint project of the University of California, Los Angeles, and the Jet Propulsion Laboratory/California
Institute of Technology, and NEOWISE, which is a project of the Jet Propulsion Laboratory/California Institute of Technology.
WISE and NEOWISE are funded by the National Aeronautics and Space Administration.
Funding for SDSS-III has been provided by the Alfred P. Sloan Foundation, the Participating Institutions,
the National Science Foundation, and the U.S. Department of Energy Office of Science.
The SDSS-III web site is \url{http://www.sdss3.org/}.
The FIRST web site is \url{http://sundog.stsci.edu/}.
The NVSS web site is \url{http://www.cv.nrao.edu/nvss/}.
G.C. expresses his deepest thanks to the Chinese Academy of Sciences for
the award of a Visiting Professorship for Senior International Scientists. Grant No. 2010T2J19.
}
\normalsize
\end{acknowledgements}

\end{document}